\documentclass[fleqn,twoside]{article}
\usepackage{espcrc2}

\def\lsim{\raise0.3ex\hbox{$\;<$\kern-0.75em\raise-1.1ex
\hbox{$\sim\;$}}}
\def\gsim{\raise0.3ex\hbox{$\;>$\kern-0.75em\raise-1.1ex
\hbox{$\sim\;$}}}

\usepackage{graphicx}
\usepackage[figuresright]{rotating}

\newcommand{\AmS}{{\protect\the\textfont2
  A\kern-.1667em\lower.5ex\hbox{M}\kern-.125emS}}

\title{
Resolving the octant $\theta_{23}$ degeneracy by neutrino 
oscillation experiments
\thanks{Talk presented at NOW2006, 
Otrant, Lecce, Italy, September 9-16, 2006, 
based on ~\cite{theta23,t2kk2}}
}
\author{Hiroshi Nunokawa
\address{
Departamento de F\'{\i}sica, Pontif{\'\i}cia Universidade Cat{\'o}lica 
do Rio de Janeiro, C. P. 38071, 22452-970, Rio de Janeiro, Brazil}
}
       
\begin{document}

\begin{abstract}
We discuss how and to what extent the degeneracy associated 
with $\theta_{23}$, if it's not maximal, 
can be resolved by future oscillation 
experiments which utilize conventional neutrino beam from 
accelerator and/or reactor neutrinos.  
\vspace{1pc}
\end{abstract}

\maketitle

\section{Introduction}
Physics of neutrinos will soon be entering into 
the era of precision physics. 
Proposed projects such as T2K~\cite{T2K} 
and NO$\nu$A~\cite{NOvA} experiment will be able to 
measure the mixing parameters responsible for atmospheric 
neutrino oscillation,
$\sin^22 \theta_{23}$ and $\Delta m^2_{23}$ with 
a few percent uncertainty.
Moreover, these experiments can probe
$\sin^22 \theta_{13}$ down to $\sim 10^{-2}$ or smaller. 
If non-zero $\theta_{13}$ will be established, 
the future phases of these experiments 
will be aimed to determine the CP violating phase 
as well as the mass hierarchy~\cite{T2K,NOvA} 
(see also \cite{nova_t2k}). 

It has been known that in order to perform precise determinations 
of neutrino mixing parameters including the CP phase 
and mass hierarchy, 
one must confront with the problem of 
so called parameter degeneracy. 
There are 3 types of such degeneracy, 
octant~\cite{octant}, intrinsic~\cite{Burguet-C} 
and sign $\Delta m^2$~\cite{MNjhep01} degeneracy. 
Here we focus only on the octant degeneracy, 
which is decoupled from the other ones with a 
good approximation in our experimental set up.  
We consider only experiments based on the 
conventional neutrino beam from accelerator 
and reactor neutrinos. 

Suppose that 
$\theta_{23}$ is different from $\pi/4$. 
Disappearance $\nu_\mu \to \nu_\mu$ experiment can 
determine quite precisely the value of $\sin^2 2\theta_{23}$ 
with 1\% uncertainty~\cite{T2K} but
this does not allow us to know in which octant 
$\theta_{23}$ lives, and 
leads, at first approximation, ignoring 
$\theta_{13}$, to the following 2 degenerate solutions,  
$\sin^2 \theta_{23} = 
 \left[ 1 \pm \sqrt{1-\sin^2 2\theta_{23}}\right]/2$. 
For example, $\sin^2\theta_{23} = 0.4$ or 0.6 (0.45 or 0.55) if 
$\sin^22\theta_{23} = 0.96$ (0.99). 
How and to what extent 
this can be resolved is the topics of this talk. 

\section{Combining Accelerator and Reactor}

Let us first discuss the possibility to resolve this 
degeneracy by combining accelerator and reactor neutrinos~\cite{theta23},
based on the suggestion in ~\cite{reactor_theta13}. 
As a concrete example, we consider the second phase of 
the T2K experiment~\cite{T2K} with upgraded beam power of 4 MW and 
Hyper-Kamiokande (HK) detector, and
high statistics second generation reactor experiment, e.g. Angra
project~\cite{Angra}. 
We fix the mixing parameters relevant for solar neutrinos 
as $\Delta m^2_{21} = 8.0\times 10^{-5}$ eV$^2$ and 
$\sin^2 \theta_{12}$ = 0.31, and the atmospheric 
$\Delta m^2_{23} = 2.5\times 10^{-3}$ eV$^2$. 
The exposures for accelerator are assumed to be 2 (6) years of neutrino 
(anti-neutrino) running with HK whose fiducial 
volume is 0.54 Mt, whereas for the reactor we assume 
an exposure of 10 GW$\cdot$kt$\cdot$yr, which is 
defined as a product of reactor power, detector volume and running time.  

In Fig. 1 we show the process of how the octant degeneracy can 
be resolved by combining the results from accelerator and reactor. 
Figs. 1(a) and (b) show that by combining the $\nu_\mu \to \nu_\mu$ 
and $\nu_\mu \to \nu_e$ modes from accelerator, 
we can get 2 separated allowed regions where one of them 
corresponds to the fake solution. Then if we add the result
of the measurement of $\theta_{13}$ form reactor experiment, 
which is shown in Fig. 1(c), we can 
eliminate the fake solution and end up with the true one
as shown in Fig. 1(d).

\newpage
\vglue -0.9cm
\hglue -0.5cm
\includegraphics[width=0.55\textwidth]{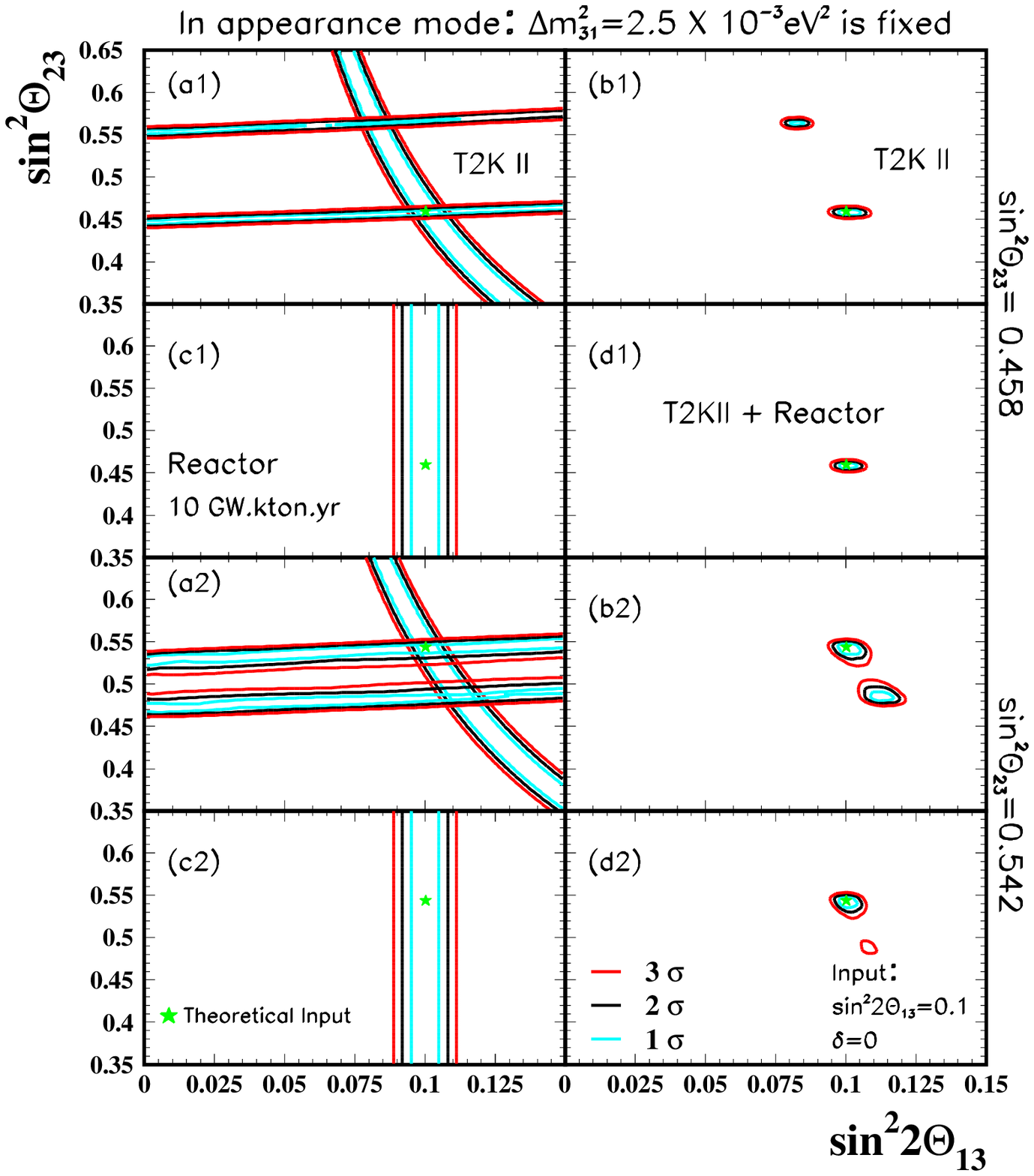}
\vglue -0.5cm
\noindent
Figure 1.
(a) Region allowed by the disappearance (2 approximately 
horizontal bands) and by the appearance (the other curved band)
accelerator measurement.
(b) The regions that remain allowed when results shown 
in (a) are combined.
(c) The regions  allowed by reactor measurement.
(d) The regions allowed after combining all the results 
from accelerator and the reactor experiments. 
Taken from ~\cite{theta23}. 
\vglue 0.30cm


We show in Fig.~2 the parameter region 
of $\sin^2 2\theta_{13}$ and $\sin^2 \theta_{23}$
where the octant degeneracy can be resolved. 
The degeneracy can be resolved for
$\sin^2 2\theta_{23} \lsim$ 0.96 at 2 $\sigma$ CL 
for $\sin^2 2\theta_{13} \gsim$ 0.05 (see upper panel). 
We confirmed the strong dependence of sensitivity on $\theta_{13}$ 
expected in \cite{reactor_theta13}. 

\section{Accelerator based 2 detector method}

Next we discuss another possibility based on~\cite{t2kk1}. 
In the second phase of T2K experiment (HK with 4 MW beam), 
in order to improve the sensitivity to the mass hierarchy determination, 
it was suggested to place the second identical 
detector at Korea with baseline of 1050 km,
in addition to the one at Kamioka. 
Here we assume 2 detectors are not only identical 
but also receive the neutrino beam with the same
energy spectrum by choosing the same off axis angle
(2.5 degree in this case). 
It was shown that this experimental 
set up can also resolve the octant degeneracy~\cite{t2kk2}. 

\vglue -1.3cm
\includegraphics[width=0.545\textwidth]{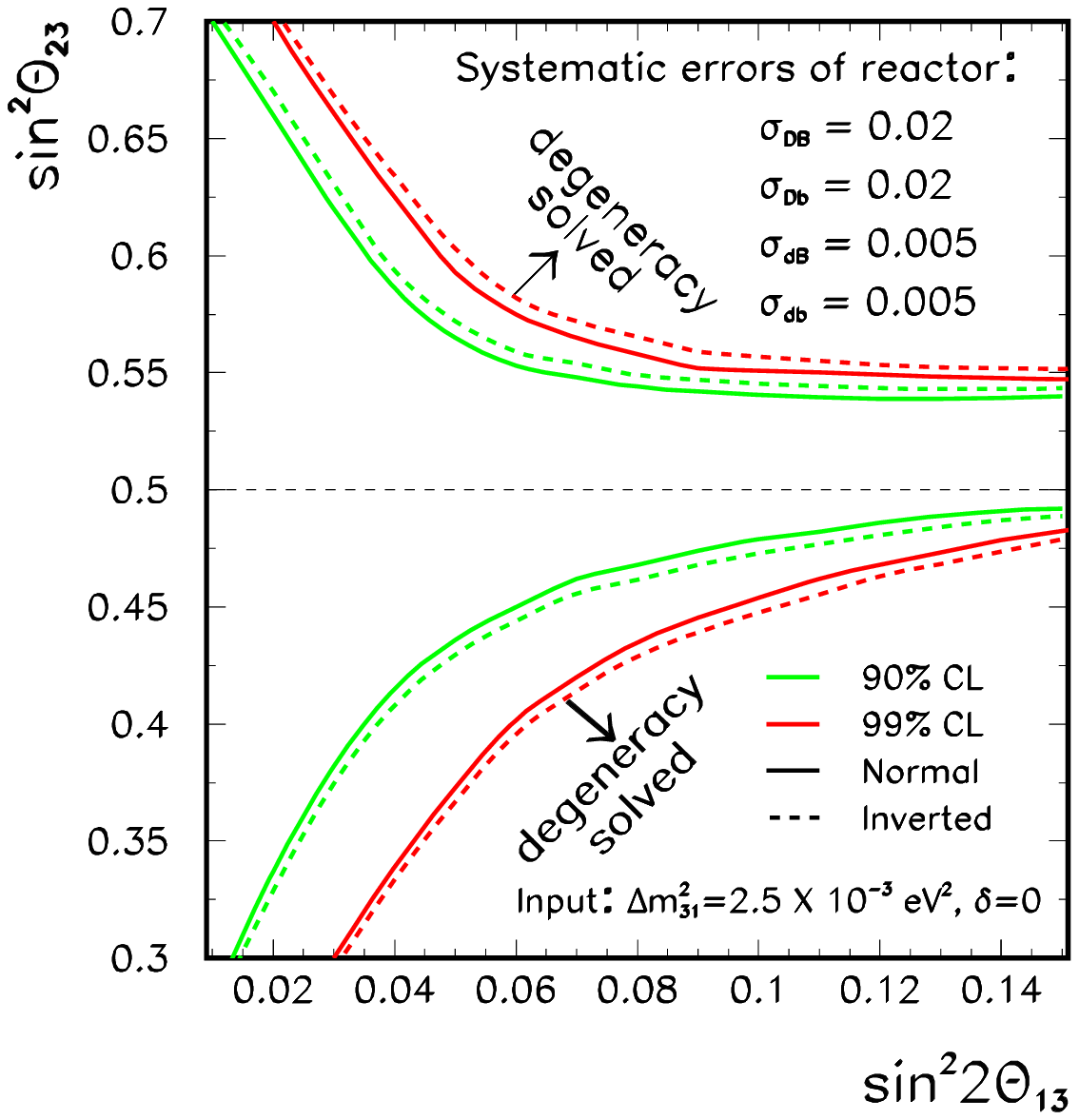}
\vglue -3.6cm
\includegraphics[width=0.545\textwidth]{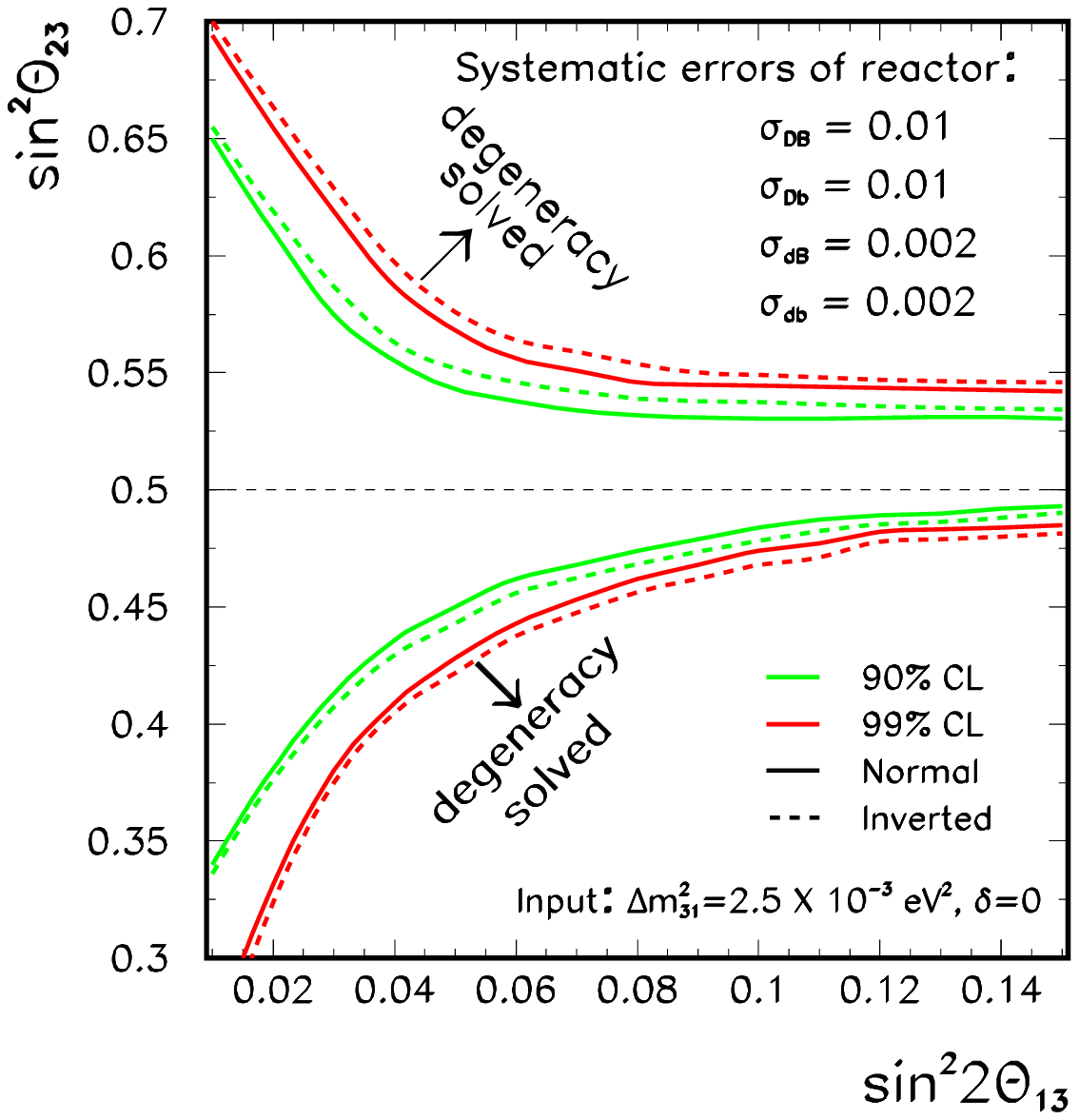}
\vglue -2.1cm
\noindent 
Figure 2. Region of parameters where $\theta_{23}$ degeneracy 
can be resolved. Upper (lower) panel corresponds to 
conservative (optimistic) choice of the systematic uncertainties.
Taken from \cite{theta23}. 
\label{fig:reactor-sensitivity2}
\vglue 0.3cm

The idea is to detect the ``solar'' term defined as
$\cos^2\theta_{23} \sin^2 2  \theta_{12} 
(\Delta m^2_{21}L/4E)^2$, which is 
possible if the detector is placed far enough. 
Since this term is proportional to $\cos^2\theta_{23}$,  
in principle, one can tell in which octant $\theta_{23}$ lies.
In Fig. 3, we show the expected number of muon and electron 
events as a function 
of reconstructed neutrino energy at Kamioka and Korea 
for $\sin^2\theta_{23} = 0.4$ and 0.6. 
We observe that at Korea for the electron events, 
appreciable difference exist between the data corresponding 
to these 2 degenerate solutions, which can tell us the correct
octant.

In Fig. 4 we show the region of parameters
where the octant degeneracy can be resolved by the 2 detector method. 
Let us note that resolving 
power (sensitivity) does not depend 
much on the 
value of $\theta_{13}$ which is very different 
from the
\\
\vglue -1cm
\newpage
\vglue -0.4cm
\hglue -0.4cm
\includegraphics[width=0.44\textwidth]{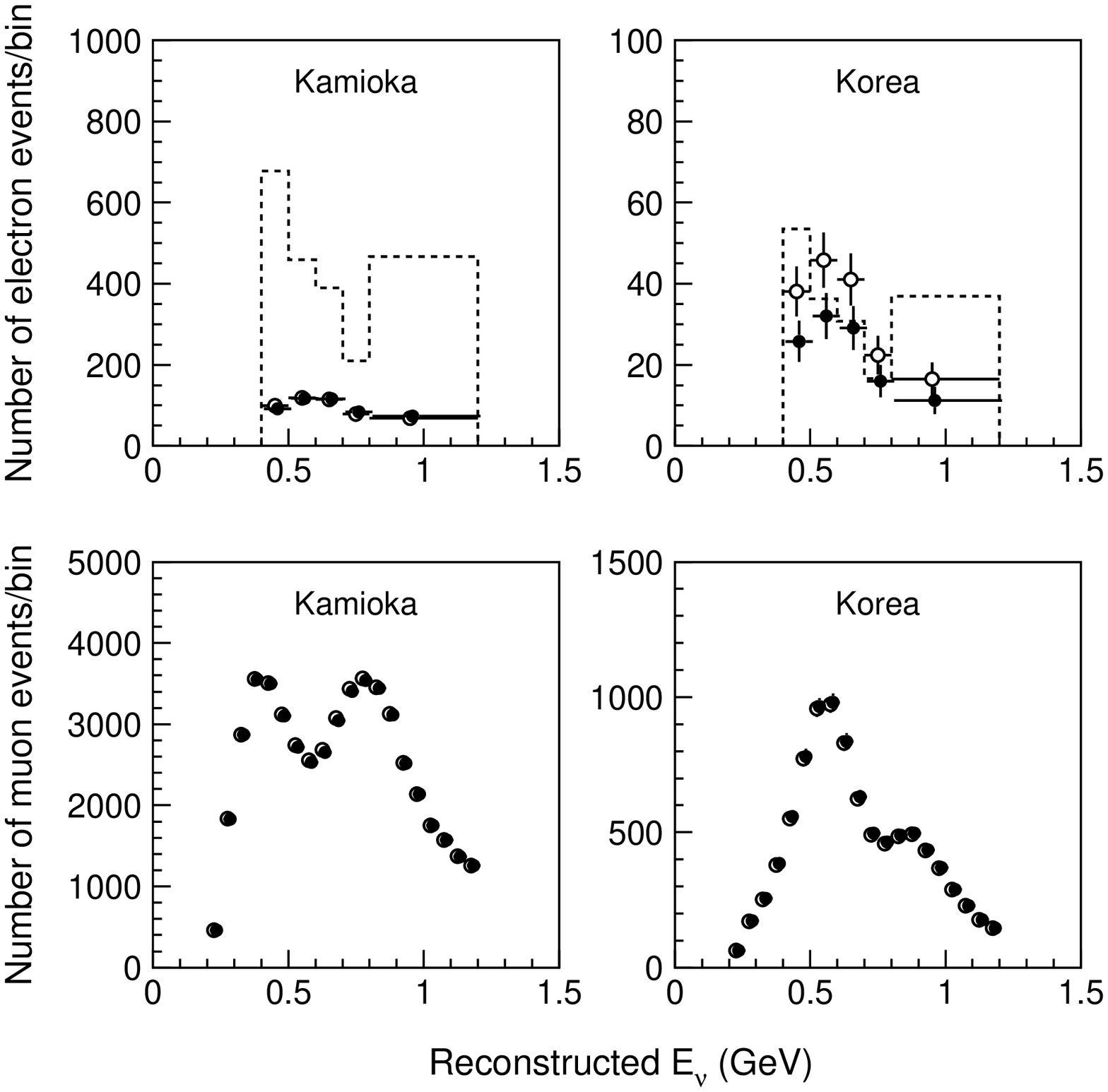}
Figure 3. Expected number of electron (upper panels) and 
muon (lower panels) events as a function of 
the reconstructed neutrino energy for $\sin^2 \theta_{23} =$0.40 
(open circles) and 0.60 (filled circles).  
Taken from \cite{t2kk2}. 
\label{fig:t2kk-strategy}
\vglue 0.6cm

\vglue -0.4cm
\hglue -0.6cm
\includegraphics[width=0.47\textwidth]{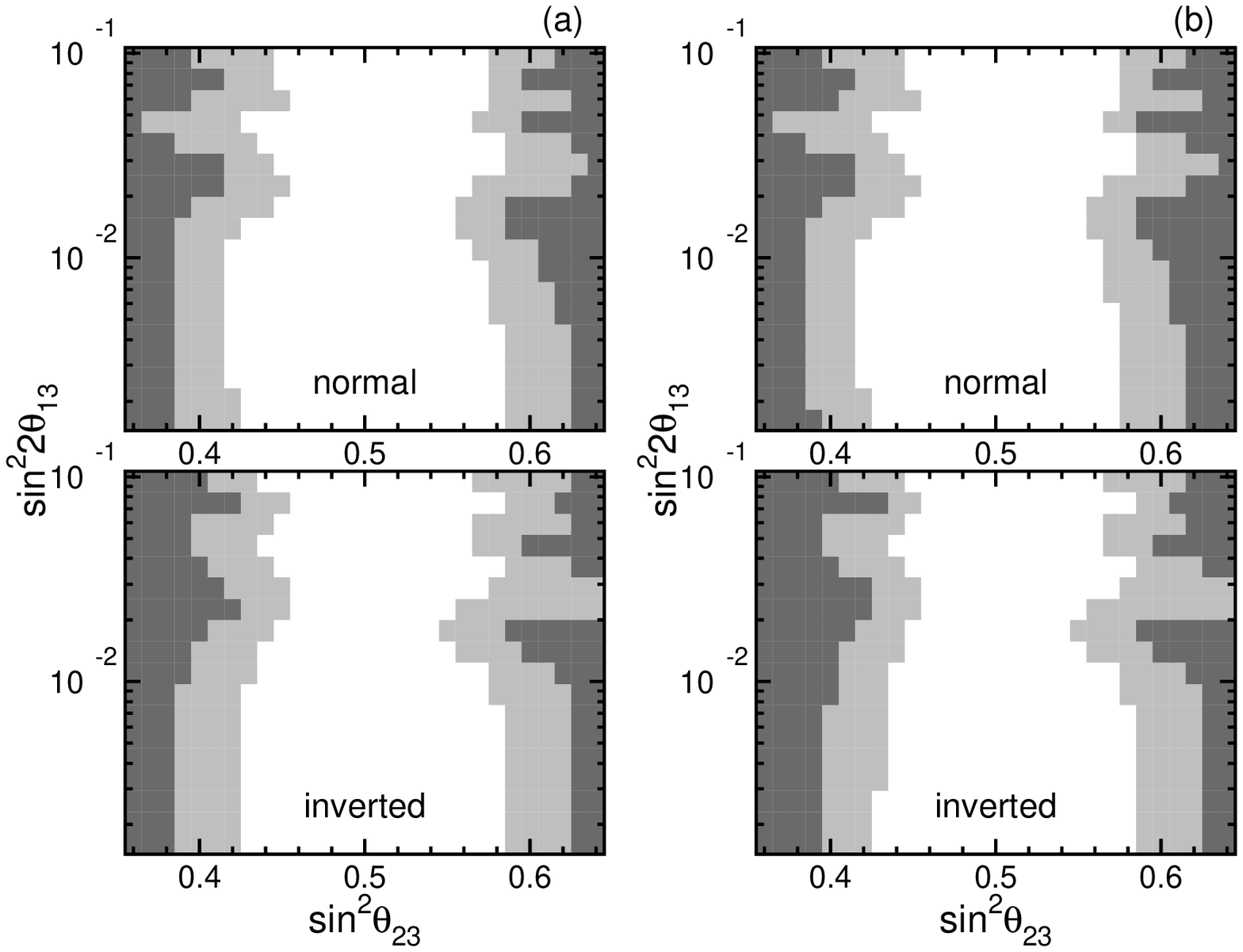}
Figure 4. 
Region of parameters where $\theta_{23}$ degeneracy 
can be resolved. Light gray (dark gray) area  
corresponds to 2 (3) $\sigma$ significance. 
0.27~Mton detectors both in Kamioka and Korea 
and 4 years running with neutrino beam and another 4 years 
with anti-neutrino beam are assumed.
In left (right) 2 panels, the sensitivity is defined so that the experiment 
can resolve the octant degeneracy 
for any (half) values of the CP phase $\delta$. 
Taken from \cite{t2kk2}. 
\vglue 0.2cm
%

\noindent
previous case shown in Fig. 2 where the sensitivity 
is significantly worse (better) when $\theta_{13}$ is small (large). 
We conclude that in this method, 
the octant degeneracy can be resolved for 
$\sin^2 2\theta_{23} \lsim$ 0.97 at 2 $\sigma$ CL 
even for very small values of $\theta_{13}$.

\section{Summary}
The expected sensitivities by the 2 methods we discussed 
show very different dependence on $\theta_{13}$. 
If $\theta_{13}$ is larger, close to the present limit, 
the method of combining accelerator and reactor
would give better sensitivity whereas for smaller $\theta_{13}$, 
the accelerator based 2 detector methods would be better,
and therefore, these 2 methods are complementary to each other. 
See Refs.~\cite{theta23,t2kk2} for details. 

\vglue 0.2cm
\noindent
{\bf Acknowledgment.} The author would like to thank
K. Hiraide, T. Kajita, H. Minakata, T. Nakaya, 
S. Nakayama, H. Sugiyama, W.~J.~C.~ Teves and 
R. Zuaknovich Funchal for collaborations. 
The author acknowledges also CNPq and FAPERJ 
for financial support. 

\vglue -0.5cm


\begin{thebibliography}{9}

\bibitem{theta23}
 K.~Hiraide {\it et al.}, 
  Phys.\ Rev.\ D {\bf 73} (2006) 093008
  [arXiv:hep-ph/0601258].

\bibitem{t2kk2}
  T.~Kajita {\it et al.}, 
  arXiv:hep-ph/0609286, 
to be published in   Phys.\ Rev.\ D. 

\bibitem {T2K}
Y.~Itow {\it et al.}, arXiv:hep-ex/0106019.\\
For an updated version, see: http://neutrino.
kek.jp/jhfnu/loi/loi.v2.030528.pdf

\bibitem {NOvA}
 D.~S.~Ayres {\it et al.}  [NOvA Collaboration],
  arXiv:hep-ex/0503053.

\bibitem{nova_t2k}
O.~Mena {\it et al.},
arXiv:hep-ph/0609011.


\bibitem{octant}
G.~Fogli and E.~Lisi, Phys.\ Rev.\ {\bf D54} (1996) 3667 
[arXiv:hep-ph/9604415].


\bibitem{Burguet-C}
J.~Burguet-Castell {\it et al.}, 
Nucl.\ Phys.\ B {\bf 608} (2001) 301
[arXiv:hep-ph/0103258].

\bibitem{MNjhep01}
H.~Minakata and H.~Nunokawa,
JHEP {\bf 0110} (2001) 001 [arXiv:hep-ph/0108085]. 



\bibitem {Angra}
J.~C.~Anjos {\it et al.},
 Nucl.\ Phys.\ Proc.\ Suppl.\  {\bf 155} (2006) 231
 [arXiv:hep-ex/0511059].

\bibitem{reactor_theta13}
H.~Minakata {\it et al.},
Phys.\ Rev.\ D {\bf 68} (2003) 033017
[Erratum-ibid.\ D {\bf 70} (2004) 059901]
[arXiv:hep-ph/0211111].


\bibitem{t2kk1}

M.~Ishitsuka {\it et al.}, 
  Phys.\ Rev.\ D {\bf 72} (2005) 033003
  [arXiv:hep-ph/0504026].

\end{thebibliography}
\end{document}